\documentclass[a4paper,12pt]{article}
\usepackage{amssymb}
\usepackage{amsmath}
\begin{document}

\title{ If the spin of gravitational field depends on spacetime dimension}
\author{K. Kaviani $^{1}$\footnote{kamran.kaviani@gmail.com}  and F. Atyabi $^{2}$\footnote{farzaneh.atyabi@gmail.com} \\
\footnotesize {$^{1,2}$ Department of Physics, Alzahra University, Tehran, Iran} \\}
\date{}
\maketitle

\newtheorem{The}{Theorem}
\newtheorem{lem}{Lemma}
\newtheorem{prop}{Proposition}
\newtheorem{hyp}{Hypothesis}
\newtheorem{rem}{Remark}
\markright{}
 \begin{abstract}
In this manuscript a different perspective on gravitational field has been proposed, suggesting different features of gravity depending on spacetime dimension, which can explain preventing the formation of singularity inside blackholes and also suggest interpreting spin-$1$ fields emerged in theories with dimensional reduction mechanism, as  different aspects of gravitational field.
\end{abstract}

\section{Introduction} 
Uncovering mysterious feature of quantum gravity and also unification of all fundamental interactions, which is not independent of the quest for a quantum theory of gravity, have received a great deal of attention from physicists. The persuit of these goals has been realized through some approaches with different insights. One of the most prominent candidates is loop quantum gravity(LQG)\cite{Rovelli,Thiemann} which makes attempts to quantize gravity introducing a spin foam structure for spacetime without using a background spacetime. Another one is string theory\cite{Polchinski1,Polchinski2}, with the aim of incorporating gravity with other fundamental forces, based on the insights of Kaluza-Klein theory which was one of the first attempts to create a unified field theory.

In this manuscript we have a proposal introducing a different feature of gravitational field depending on spacetime dimension which can explain some inconsistencies in the mentioned theories and also absence of singularity inside blackholes.
\section{Dimensional reduction and emerging different features of gravity}
 Although we live in a $4D$-world, but we think that in this $4D$-space, some regions might be found with effectively less than four dimensions. As an example, at the center of a blackhole we may find such places. On the other hand some theoretical models possesses dimensional reduction mechanism, such as Kaluza-Klein and ADS/CFT theories which we mention briefly their mechanisms.
 \subsection{The Kaluza-Klein theory}
In prototype Kaluza-Klein theory \cite{Collins,Coquereaux}, general relativity is extended to a five dimensional spacetime. By compactification of one of the spatial dimensions, an ordinary theory of gravity in four dimensions together with a Maxwellian theory of electromagnetism can be achieved. However this theory encounters some unsatisfactory features, spechially as far as matter dynamic is concered such that deviation of electron mass and electric charge ratio from experimental data is egregious and appearing electromagnetism force with the same order of magnitude of gravity force can not be justified.
\subsection{AdS/CFT correspondence}
A somehow similar procedure can be seen in AdS/CFT correspondence\cite{Maldacena,Gubser,Witten,Hoker}, which was discovered in the context of string theory and is called gauge/gravity duality which relates a theory with gravity in a $d$ dimensional anti-de sitter space to a local field theory without gravity in a $d-1$ dimensional space. As a basic AdS-CFT scenario one can refer to equivalence between a gravitational theory in $AdS_{5}\times S^{5}$ \footnote{Five-dimensional Anti-de Sitter space times a compact five sphere}
 and a four-dimensional super Yang-Mills theory. In the equivalence, fundamental coupling constant of the theory, string coupling $g_{s}$ and the Yang-Mills coupling $g_{YM}$, are related by $g^{2}_{YM}=4\pi g_{s}$. The duality states that information of quantum gravity in the five-dimensional space can be recaptured by the Yang-Mills field on boundary of the space and that is why the duality is referred to as a holographic duality. 
 \subsection{Proposal}
However these theories are considered in different frameworks and the emerged fields after dimensional reduction are usually interpreted as completely new fields, but by considering the similarity in the mentioned aspects of these theories and the Yang-Mills coupling predicted by the theories, one might conclude that the emerged fields can not describe fundamental interactions except gravity.
If one can postulate dependence of gravitational field spin on spacetime dimension,  these new fields can be interpreted as different features of gravity.
As we know spin of gravitational field is two. We propose that it is just the case for four-dimensional space and in general might be half of the spacetime dimension. This proposal provides possibility to explain some precious physical features which will be explained in the following.

\section{physical features}
 \subsection*{$\bullet$ Explanation for Yang-Mills coupling in the theories with dimensional reduction}
By considering Yang-Mills coupling predicted by Kaluza-Klein theory and the gauge/gravity duality through dimensional reduction, we conclude that the emerged spin-$1$ fields can not be considered as fields with physical interpretation, since in our $4D$-world we have not experienced physical interactions due to the existance of such fields.

According to our proposal, reduction from higher to lower dimensional space is associated with emerging some different features of gravity and the appeared spin-$1$ fields can be interpreted as gravitational fields such that existance of these fields can just be expected to examine in a two dimensional space not in our $4D$-world.
 \subsection*{$\bullet$ Explanation for resolving blackhole singularity}
Curvature singularity is predicted by Einstein's general relativity for gravitation. Such a prediction indicates that the geometrical description offered by general relativity fails to give a consistent picture at spacetime singularities. The spacetime sigularities must be resolved with a theory of gravity in which blackhole singularity and Bigbang singularity are controlled by quantum effects \cite{Rovelli,Ashtekar,Bojowald,Modesto,Ashtekar2}.\\
   Now consider a static blackhole. The shwarzschild metric, $ds^2=-(1-\frac{2GM}{r})dt^2+(1-\frac{2GM}{r})^{-1}dr^2+r^2d\Omega^2$, describes the spacetime curvature around the blackhole with a curvature singularity located at $r=0$. As $r\rightarrow0$ one obtains the metric independent of the angular coordinate $\Omega$. So in this limit we effectively have two dimensional spacetime and according to the proposal, the spin of gravitational field might change to one, so gravitational force law changes such that one might see a different aspect of gravity as a repulsive force in the same order of magnitude of attractive force arising from the spin 2 field, therefore forming a singularity is prevented according to the proposal.\footnote{A similar procedure is shown in reference \cite{Smoller}, where a Yang-Mills gauge theory coupled to Einstein gravity, can make a repulsive force to balance gravitational attractive force preventing spacetime singularity.}
\section*{Acknowledgement}
We would like to thank Prof. Valeriy Dvoeglazov for his fruitful discussion.

\end{document}